\documentclass[letterpaper]{jpconf}
\usepackage{graphicx}
\usepackage{epstopdf}
\begin{document}
\title{Quantum criticality and black holes}

\author{Subir Sachdev and Markus M\"uller}

\address{Department of Physics, Harvard University, Cambridge MA 02138, USA}


\begin{abstract}
Many condensed matter experiments explore the finite temperature dynamics
of systems near quantum critical points. Often, there are no well-defined quasiparticle
excitations, and so quantum kinetic equations do not describe
the transport properties completely. The theory shows that the transport co-efficients are not proportional to a mean
free scattering time (as is the case in the Boltzmann theory of quasiparticles),
but are completely determined by the absolute temperature and by equilibrium
thermodynamic observables. Recently, explicit solutions of this quantum critical
dynamics have become possible via the AdS/CFT duality discovered 
in string theory. This shows that the quantum critical theory provides a holographic
description of the quantum theory of black holes in a negatively curved anti-de Sitter
space, and relates its transport co-efficients to properties of the Hawking radiation
from the black hole. We review how insights from this connection have led to new
results for experimental systems: ({\em i\/}) the vicinity of the superfluid-insulator
transition in the presence of an applied magnetic field, and its possible application
to measurements of the Nernst effect in the cuprates, ({\em ii\/}) the magnetohydrodynamics
of the plasma of Dirac electrons in graphene and the prediction of a hydrodynamic
cyclotron resonance.
\\
~\\
{\bf Plenary talk at the 25th International Conference on Low Temperature Physics,
Amsterdam, 6-13 Aug 2008.}
\end{abstract}

\section{Introduction}

A major focus of research in condensed matter physics is on systems which are not
described by the familiar paradigms of order parameters and quasiparticles. In states
with broken symmetry, we focus on the order parameter which characterizes the broken symmetry,
and work with effective classical equations of motion for the order parameter. In states with long-lived
quasiparticle excitations, we write down quantum transport equations ({\em e.g.\/} via the Keldysh formulation)
for the quasiparticles and deduce a variety of transport properties. However, it has become clear
in recent years that a number of interesting correlated electron materials do not fit easily into either
of these paradigms. One promising and popular approach for describing these systems is to 
exploit their frequent proximity to quantum phase transitions \cite{ssbook}, and so use a description in terms
of ``quantum criticality''. This term refers to dynamics and transport in the non-zero temperature ($T$) 
quantum critical region which
spreads out from the $T=0$ quantum critical point or phase (see Fig~\ref{qcfig} below). 
Because there are no well-defined quasiparticles,
and the excitations are strongly interacting, the description of quantum criticality is a challenging problem.

It is useful to begin our discussion of quantum criticality by recalling a well-understood 
example of non-quasiparticle and non-order parameter dynamics: the Tomonaga-Luttinger (TL)
liquid \cite{giam}. This is a compressible quantum state of many fermion or many bosons systems confined to move
in one spatial dimension. Electronic quasiparticles are not well defined in this state, but there is a well-developed
theory of the transport properties of the TL liquid. Historically, this theory evolved over
several decades of research on quantum many body systems in one dimension. Key in the historical development 
were exact solutions of model Hamiltonians via the Bethe Ansatz. Insights gained from the structure of excitations
in the Bethe Ansatz solutions led to a more general understanding of the low energy excitations of a generic Hamiltonian,
and a universal low energy theory of the TL liquid. Thus while the exact solutions were restricted
to artificial models, they played a key role in the development of the general theory. Of course, after the fact, with
the general theory of the TL liquid before us, we can justify it in its own terms, and largely dispense
with reference to the Bethe Ansatz solutions.

While the TL liquid does describe quantum critical states in one dimension, many
of its characteristics do not generalize to quantum criticality in higher spatial dimensions. In particular, it is not difficult
to show \cite{giam,damless,m2cft} that the correlators of all conserved currents in TL liquids show a characteristic `free particle' ballistic behavior
at all frequencies ($\omega$) and $T$. This is because a free-field description of the conserved charges
can usually be found; there are no collisions between the excitations and consequently no relaxation to
collision-dominated hydrodynamic behavior. Hydrodynamics 
emerges only after including ``irrelevant'' perturbations to the theory of the TL
liquid. As we will review below, generic strongly-coupled quantum critical points in higher dimensions behave
very differently: there are no known free field formulations, even with variables which are highly non-local in terms
of the microscopic degrees of freedom. Furthermore, relaxation to local thermal equilibrium and hydrodynamic
behavior emerges already within the universal theory of the quantum critical point.

We will focus in this paper on strongly interacting
quantum critical points in 2+1 dimensions \cite{ssbook}. Prominent examples include the transition between antiferromagnetic
and spin gap states in Mott insulators, and the superfluid-insulator transition in the lattice boson models. 
A theory for the quantum critical transport of such models has been developed \cite{damless,sondhi1,bgs}, using insights
from various weak-coupling perturbative renormalization group analyses. However, this approach was limited
to a narrow range of physical parameters: for the case of the superfluid-insulator transition, the density
of the particles was exactly equal to the commensurate density of the Mott insulator and 
there were no impurities. In this article we will review how this field has 
received an impetus from an unexpected direction: the AdS/CFT duality
of string theory \cite{MAGOO}. This duality has provided a set of exact solutions \cite{m2cft,nernst} of quantum critical transport in 2+1 dimensions.
Admittedly, the exact solutions are for rather artificial models, with a high degree of supersymmetry. However, just
as was the case for the TL liquid, these exact solutions have provided an impetus to the general theory,
and led to significant progress in formulating a more general theory of quantum critical transport, without the 
restrictions on the physical parameters noted above.

\begin{figure}
\begin{center}
\includegraphics[width=5.5in]{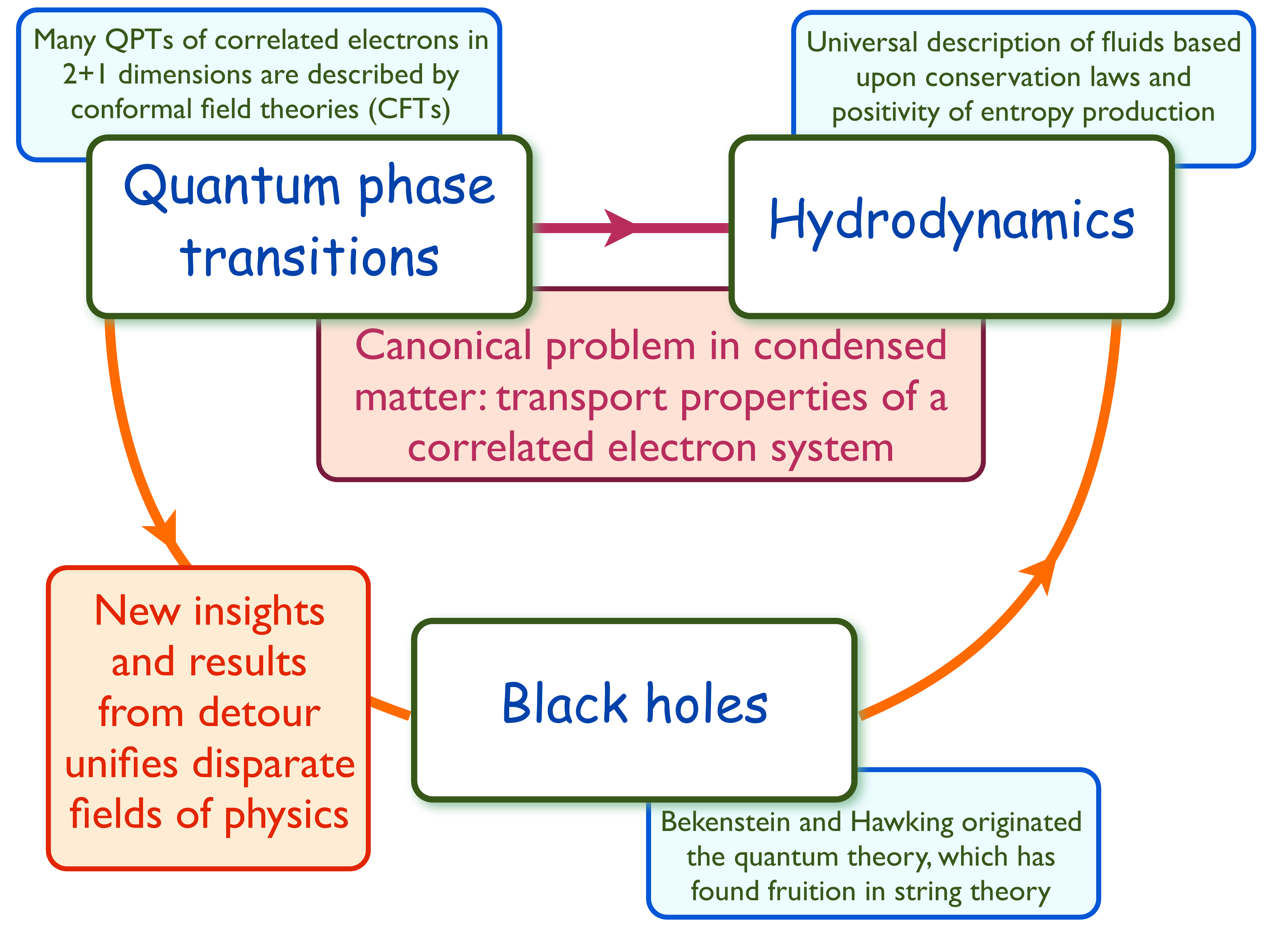}
\caption{The connections between quantum criticality and black holes described in this paper}
\label{diag}
\end{center}
\end{figure}

We have summarized the conceptual relationship between the developments in condensed matter and string theory in Fig.~\ref{diag}.
The remainder of the article will expand on connections shown in this figure. In Section~\ref{sec:qct}, we will review
the theory of quantum critical transport from a traditional condensed matter perspective. The modern developments
in black holes and string theory will be summarized in Section~\ref{sec:ads}. Our new results for quantum critical 
transport in 2+1 dimensions will be presented in Section~\ref{sec:sit}. Finally, in Section~\ref{sec:recent} we briefly
note emerging new directions of research in the application of the AdS/CFT duality to condensed matter physics.

\section{Quantum critical transport}
\label{sec:qct}

\begin{figure}
\begin{center}
\includegraphics[width=5.5in]{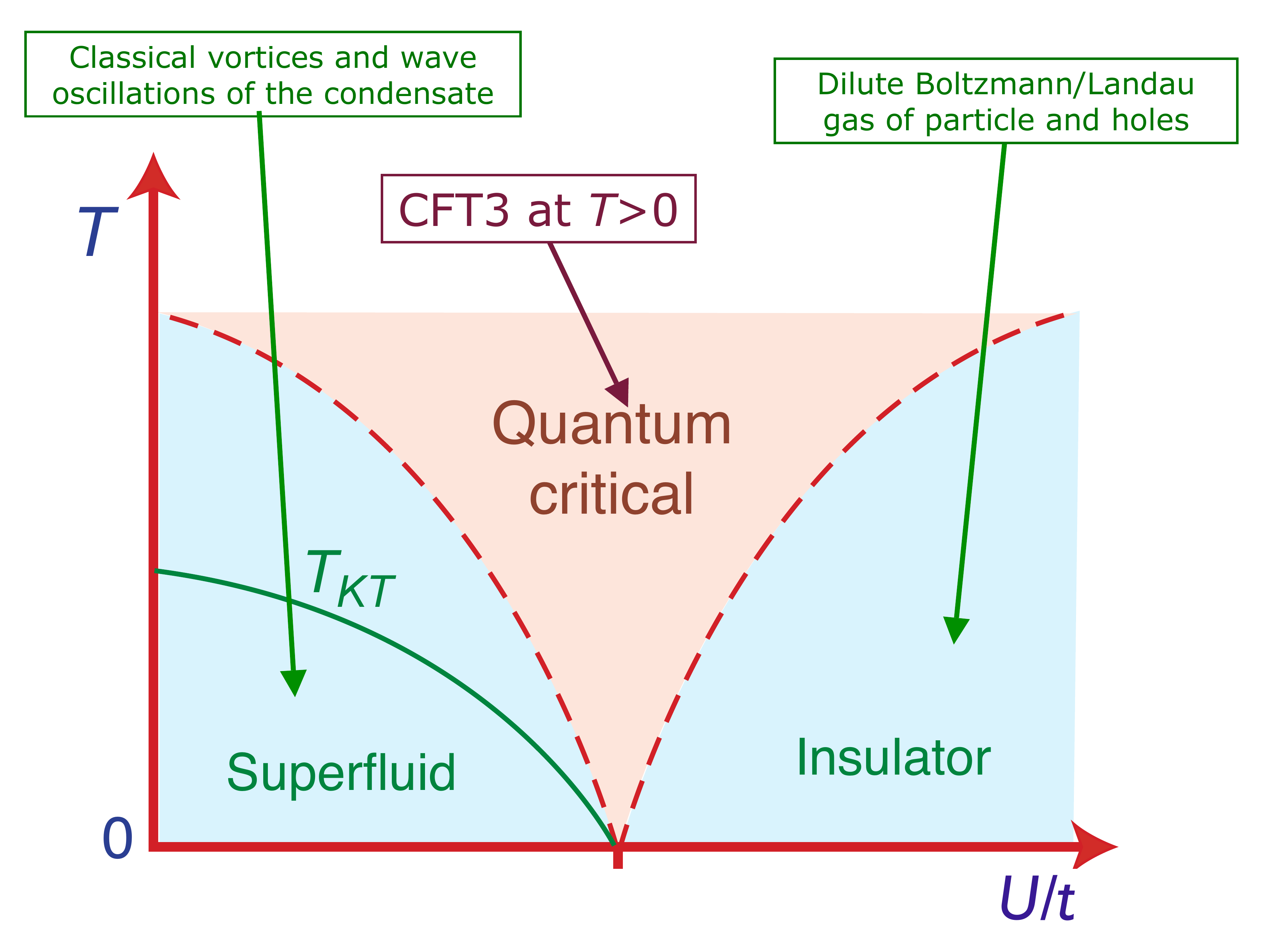}
\caption{Phase diagram of the Bose Hubbard model at integer filling. Dashed lines are crossovers, while the full
line is the Kosterlitz-Thouless transition at $T_{KT}$.}
\label{qcfig}
\end{center}
\end{figure}

Here we review the ideas presented in Ref.~\cite{damless} on transport near the superfluid-insulator
transition in two spatial dimensions. For definiteness, we will consider the simplest model \cite{fwgf} which exhibits
such a transition: the Bose Hubbard model at integer filling. As shown in Fig.~\ref{qcfig}, the transition is induced
by varying the ratio $U/t$, where $U$ is the on-site repulsion between the bosons, and $t$ is the boson
hopping matrix element.
At low $T$, and generic low $U/t$, Fig.~\ref{qcfig} shows that the system is in one of two possible regimes, both
of which have effective classical theories of their dynamics and transport. At small $U/t$, in the superfluid region and
across the Kosterlitz-Thouless transition, we can use the Gross-Pitaevski equation to describe the spin-wave and vortex fluctuations.
At large $U/t$, we have to consider the particle and hole excitations on the insulator, and these are dilute enough at low $T$
to allow a classical gas description. However, our primary interest here is in the intermediate quantum-critical regime,
where neither classical description is possible. The theory of this region involves the dynamics of the quantum critical point,
which is known to be described by a 2+1 dimensional conformal field theory (CFT3) associated with the Wilson-Fisher fixed point \cite{fwgf}.

The theory of the quantum critical region shows that the transport co-efficients, and the relaxation time to local
equilibrium, are not proportional to a mean free scattering time between the excitations, as is
the case in the Boltzmann theory of quasiparticles. Such a time would typically depend upon the interaction
strength, $\sim U$, between the particles. Rather, the system behaves like a ``perfect fluid'' in which the relaxation
time is as short as possible, and is determined universally by the absolute temperature. Indeed, it was
conjectured in Ref.~\cite{ssbook} that the relaxation time generically obeys
\begin{equation}
\tau \geq  \frac{\hbar}{k_B T} \Phi_\tau \label{tau}
\end{equation}
where $\Phi_\tau$ is a dimensionless constant of order unity determined by the precise definition of $\tau$.
The inequality in Eq.~(\ref{tau}) is saturated in the quantum critical region, while the other regimes of Fig.~\ref{qcfig}
have significantly larger $\tau$. Anticipating our discussion of black holes in Section~\ref{sec:ads}, closely
related bounds on $\tau$ have appeared recently in considerations of the semiclassical theory of gravity wave oscillations near 
black holes \cite{hod,pesci}.

The transport co-efficients of this quantum-critical perfect fluid also do not depend upon the interaction
strength, and can be connected to the fundamental constants of nature. In particular, the electrical conductivity, $\sigma$,
is given by \cite{damless,m2cft,pkk1}
\begin{equation}
\sigma_Q = \frac{e^{\ast 2}}{h} \Phi_\sigma, \label{ds}
\end{equation}
where $\Phi_\sigma$ is a universal dimensionless constant of order unity, and we have added the subscript $Q$ to emphasize that this is the conductivity for the case at integer filling,
no impurity, and at zero magnetic field. Here $e^\ast$ is the charge of the carriers: for a superfluid-insulator
transition of Cooper pairs, we have $e^\ast = 2e$, while in our application to graphene below we have $e^\ast = e$.

In addition to charge transport, we can also consider momentum transport. This was considered in the context of applications
to the quark-gluon plasma \cite{kss}; application of the analysis of Ref.~\cite{damless} shows that the viscosity, $\eta$, is given by
\begin{equation}
\frac{\eta}{s} = \frac{\hbar}{k_B} \Phi_\eta, \label{eta}
\end{equation}
where $s$ is the entropy density, and again $\Phi_\eta$ is a universal constant of order unity. The entropy density itself obeys \cite{ssbook}
\begin{equation}
s = \frac{k_B^3 T^2}{\hbar^2 v^2} \Phi_s, \label{s}
\end{equation}
where $v$ is a characteristic velocity of the CFT3, much smaller than the physical velocity of light in condensed matter realizations.
The universal constant $\Phi_s$ is the analog of the `central charge' of CFT2s in 1+1 dimension \cite{pkk2}.

The values of the universal constants $\Phi_s$, $\Phi_\tau$, and $\Phi_\sigma$ have been determined \cite{ssbook,damless,sspoly} to 
leading order in expansions
in $1/N$ (where $N$ is the number of order parameter components) and in $3-d$ (where $d$ is the spatial dimensionality).
However, both expansions for $\Phi_{\tau,\sigma}$ are neither straightforward nor rigorous, and 
require a physically motivated resummation of the
bare perturbative expansion to all orders. It would therefore be valuable to have exact solutions of quantum critical
transport where the above results can be tested, and we turn to such solutions in the section below.

\section{Black holes and the AdS/CFT correspondence}
\label{sec:ads}

We now turn to a seemingly distinct subject: the quantum theory of black holes. 

Soon after Einstein's formulation of the theory of general relativity, Schwarzschild discovered its black hole solutions:
a dense object of mass $M$ was surrounded by a horizon of radius $R = 2 G M/c^2$ which enclosed a region causally disconnected
from the rest of the universe (here $c$ is the actual velocity of light). 
The next result of relevance to us appeared in 1973, when Bekenstein \cite{beck} postulated a remarkable
connection between the properties of black holes and the laws of thermodynamics. In particular, he argued that each black hole
carried an entropy 
\begin{equation}
\frac{S}{A} = \frac{k_B}{4 \ell_P^2},  \label{sb}
\end{equation}
where $A = 4 \pi R^2$ is the area of the horizon and $\ell_P = \sqrt{G \hbar/c^3}$ is the Planck length 
(the precise numerical
factor of 1/4 appeared in the subsequent work of Hawking); thus each Planck length square on the horizon has roughly one
qubit degree of freedom. Bekenstein was partly motivated by the similarity between
the law $d A \geq 0$, which is a consequence of the dynamics of the black holes in the theory of general relativity,
and the second law of thermodynamics, $d S \geq 0$. He also presented physical arguments that without such a relation
between $S$ and $A$, we could violate the second-law of thermodynamics simply by throwing matter into a black hole.

Bekenstein's proposal was put on a firmer footing by Hawking \cite{hawking}, who considered the quantum fluctuations of fields in the vicinity
of the horizon of a black hole. He found the horizon also had a temperature, $T = \hbar^2/(8 \pi k_B M \ell_p^2)$, and emitted black body
radiation at this temperature. Using this temperature and the expression in Eq.~(\ref{sb}), we can express $S$ in terms
of $T$ (eliminating $M$ and $R$), and find $S \sim 1/T^2$. Consequently, the specific heat of a black hole is negative,
and it is thermodynamically unstable: its eventual fate is evaporation into Hawking radiation.

Recent advances in string theory have found a remarkable and explicit realization of the extension of the Bekenstein-Hawking
results to black holes in an anti-de Sitter (AdS) universe. This is a homogeneous,
negatively curved spacetime and black holes in such a universe can be thermodynamically stable (as we will see below).
The complete solution of certain supersymmetric theories of quantum
gravity in AdS spaces are now available. The central feature of the theory is the equivalence between the
quantum theory of gravity in AdS4 and a CFT3 living in a flat 2+1 dimensional spacetime: this is the famous
AdS/CFT correspondence \cite{MAGOO}. This correspondence allows computation of the black
hole entropy via its identification with the entropy of the CFT3. In a number of cases, the
result has been verified to be in precise agreement with the Bekenstein-Hawking result
in the appropriate semiclassical limit, after the $T$ of the CFT is identified with the Hawking temperature
of the black hole.

One benefit of these advances is that Bekenstein's connection between black holes and thermodynamics
can now be extended to dynamical properties \cite{kss}. We can thus relate the dynamics of gravity and matter near a black
hole to the dynamic correlations of a CFT3. In particular, transport co-efficients of the CFT3, such as $\sigma$ and $\eta$,
can be related to properties of waves propagating in the presence of a black hole in an asymptotically AdS space
(the AdS-Schwarzschild metric). The dissipation and irreversibility associated with these transport co-efficients emerges
naturally from the irreversibility of matter falling past the horizon of the black hole. Indeed, even a simple saddlepoint
treatment of the gravity theory already yields a CFT dynamics which has non-zero and non-singular transport
co-efficients, has positive entropy production, and relaxes to local thermal equilibrium. These are the features which make
the AdS/CFT correspondence so attractive: they are unprecedented in previous mean-field theories
of quantum many body systems.

We now give a brief account of the correspondence between the gravity theory on AdS4 space and the CFT3, a topic which
is extensively covered in the string theory literature. We consider physics on AdS4 space with metric
\begin{equation}
ds^2 = \frac{u^2}{L^2} (- c^2 dt^2 + dx^2 + dy^2) + L^2 \frac{du^2}{u^2} \label{ads}
\end{equation}
Here $(t,x,y)$ define the 2+1 dimensional spacetime on which the CFT3 will reside, $u$ is the fourth co-ordinate of AdS4, and $L$
is the radius of curvature. Note that here we are using the symbol $c$ (and not $v$) for the characteristic velocity of the CFT3 to connect
to the theories of black holes.
This metric is a solution of Einstein's equations with a negative cosmological constant whose value is related to $L$.
The central property of the metric in Eq.~(\ref{ads}) is that it has a large 5-dimensional isometry group
SO(3,2). This group coincides with the group of conformal symmetries in 2+1 dimensional spacetime, {\em i.e.\/} the conformal
group of CFT3s, and this is a primary reason for the AdS/CFT correspondence.  In particular, the scale invariance of the CFT3 is realized by an isometry of Eq.~(\ref{ads}) under which
\begin{equation}
(t,x,y,u) \rightarrow (t/\lambda, x/\lambda , y/\lambda ,  \lambda u).
\end{equation}
Note that while $t,x,y$ have the usual co-ordinate scaling dimension -1, $u$ has scaling dimension 1, and so scales like an energy.
Indeed, we can think of $u$ as an energy scale co-ordinate of the CFT3: $u \rightarrow \infty$ describes high energies, while
$u \rightarrow 0$ describes low energies. Of course, the physics of the unperturbed CFT3, or the infinte AdS4 space, is scale invariant, and so the properties of all $u$ are identical. 

We are invariably interested in the CFT3 when it has been perturbed by some low energy scale, such as a temperature $T>0$,
a non-zero chemical potential, or an applied magnetic field. In the dual gravity theory, such perturbations will appear
as deviations in the metric from the perfect AdS4 form at low energies {\em i.e.\/} at small $u$. We consider here the simplest
case of a non-zero $T$: this corresponds to choosing a black hole solution of Einstein's equations with a negative cosmological constant,
so that
the Hawking temperature of the black hole is $T$. Such black hole solutions are given by the AdS-Schwarzschild metric \cite{hhh1}:
\begin{equation}
ds^2 = - f(u) c^2 dt^2 + \frac{du^2}{f(u)} + \frac{u^2}{L^2} ( dx^2 + dy^2), \label{adss}
\end{equation}
where
\begin{equation}
f(u) = \frac{u^2}{L^2} - \frac{GM}{uc^2},
\end{equation}
and $M$ is the mass of the black hole. Note that (\ref{adss}) reduces to (\ref{ads}) as $u \rightarrow \infty$ {\em i.e.\/} at high energies.
The black hole horizon is at $u=u_0$ where $f(u_0) = 0$. Hawking's computation \cite{hawking} 
gives a prescription\footnote{The simplest way of computing the Hawking temperature is to analytically continue the metric
in Eq.~(\ref{adss}) to imaginary time, and to note that the resulting space is periodic in imaginary time with period $\hbar/(k_B T)$.}
for the Hawking temperature
in terms of the asymptotic behavior of the metric near $u=u_0$:
\begin{equation}
k_B T = \frac{\hbar c f' (u_0)}{4 \pi} = \frac{3 \hbar (GMc)^{1/3}}{4 \pi L^{4/3}}. \label{adst}
\end{equation}
Note that unlike the flat space case considered by Hawking, the temperature of the AdS black hole $T \sim M^{1/3}$, and so 
$T$ increases with increasing $M$; this will be important below for the thermodynamic stability of the black hole.
Also, the metric (\ref{adss}) shows that the horizon of the AdS black hole 
is actually the entire flat $x$-$y$ plane at $u=u_0$; because of this the black hole is more properly 
called a ``black brane''.  The area
of the horizon is $A= A_{xy} u_0^2/L^2$, where $A_{xy}$ is the (infinite) area of the flat $x$-$y$ plane. 
From Eq.~(\ref{sb}), the Bekenstein-Hawking entropy of the AdS black hole/brane is 
\begin{equation}
\frac{S}{A_{xy}} = \frac{k_B u_0^2}{4 \ell_P^2  L^2} = \frac{k_B G^{2/3} M^{2/3} }{4 c^{4/3} L^{2/3} \ell_P^2} \label{sxy}
\end{equation}
Eliminating $M$ between Eqs.~(\ref{adst}) and (\ref{sxy}) we see that $S \sim T^2$; so the specific heat is positive
and the black hole is thermodynamically stable.
A key point in the AdS/CFT correspondence is that the black hole entropy in Eq.~(\ref{sxy}) should be identified with
the entropy density of the CFT3 in Eq.~(\ref{s}) at a temperature given by Eq.~(\ref{adst}). Comparing these expressions
for the entropy we observe that they both scale as $S \sim T^2 \sim M^{2/3}$ and so are in agreement with each other. The numerical
prefactors also agree if we have (after noting $v=c$)
\begin{equation}
\Phi_s =  \left(\frac{2 \pi L}{3 \ell_P} \right)^2
\end{equation}
We note that the Bekenstein-Hawking entropy in Eq.~(\ref{sb}) only applies in the semiclassical gravity limit,
and so the above value of $\Phi_s$ only applies in the limit $L \gg \ell_P$. For $L \sim \ell_P$, the entropy 
would have to be computed directly from the CFT3.

The AdS/CFT correspondence has the remarkable consequence that the spacetime itself, and
its dimensionality is an ``emergent'' property. From our discussion above,
we may conclude that a CFT3 is better interpreted as ``living'' in 2+1 dimensions
when $\Phi_s \sim 1$, while some CFT3s with $\Phi_s \gg 1$ are secretly theories of gravity in 3+1
dimensions with weak quantum fluctuations.

Other important perturbations to the AdS-Schwarzschild metric have been described in the literature: a chemical potential applied
ot the CFT3 corresponds to an electric charge on the black hole, while a magnetic field turns into a magnetic charge on the 
black hole \cite{sean1}.
We will not present these metrics here, but computations \cite{nernst} on such ``dyonic'' black holes led to the results presented in Section~\ref{sec:sit}.

We conclude this section by noting the recipe \cite{MAGOO} by which dynamic correlators and transport co-efficients of the CFT3 can be computed.
Each operator in the CFT3, $\mathcal{O}$, is ``dual'' to some field, $\phi$, in the AdS4 gravity theory. For our computations
below, we note that the electrical current, $J_\mu$, of the CFT3 is dual to a U(1) gauge field, while the stress energy tensor, $T_{\mu\nu}$,
is dual to fluctuations of the metric on AdS4. The central relation between the correlations of $\mathcal{O}$ and $\phi$ is \cite{gubser0,witten}
\begin{equation}
\left\langle \exp \left( \int dx dy dt \, \mathcal{O}(x,y,t) \phi_0(x,y,t) \right) \right\rangle_{\rm CFT3} = Z_{\rm AdS4} \Bigl[
\phi (x,y,t,u\rightarrow \infty) = u^{\Delta-3} \phi_0 (x,y,t) \Bigr] \label{dual}
\end{equation}
Here $\phi_0 (x,y,t)$ is an arbitrary source field, and $\Delta$ is the scaling dimension of $\mathcal{O}$. 
So the left-hand-side of Eq.~(\ref{dual}) is the generating function of correlators
of $\mathcal{O}$ in the CFT3, a quantity of great interest to us. On the right-hand-side we have the complete partition function
of quantum gravity on AdS4 subject to a boundary condition on the fluctuating field $\phi$ as $u \rightarrow \infty$.
Note the consistency of the power of $u$ in the boundary condition with the identification of $u$ as an energy scale:
operators with smaller $\Delta$ have their dual field $\phi$ larger at smaller values of $u$ {\em i.e.\/} they are more
important in the infrared. For real time correlations at $T>0$, there are some additional analytic continuation
subtleties in applying Eq.~(\ref{dual}): these were resolved in Refs.~\cite{malda,sonherzog}.

For complete quantitative computations, the correspondence in Eq.~(\ref{dual}) is only useful when the gravity theory on
AdS4 is weakly coupled. This is known to be the case for a particular CFT3: the Yang-Mills gauge theory with $\mathcal{N}=8$
supersymmetry and a SU($N$) gauge group. In the limit $N\rightarrow \infty$, the dual theory is classical supergravity on 
AdS4. From computations on the latter theory, we can obtain precise numerical results for the Yang-Mills CFT3: the numerical 
constants in Eqs.~(\ref{ds}-\ref{s}) are given by \cite{kleb,kss,chris,m2cft}:
\begin{equation}
\Phi_\sigma = \sqrt{\frac{2 }{9} } N^{3/2}~~;~~\Phi_\eta = \frac{1}{4 \pi}~~;~~\Phi_s = \frac{8 \sqrt{2} \pi^2}{27} N^{3/2}.
\end{equation}
These are the first exact results for a strongly-coupled CFT in 2+1 dimensions.

\section{The superfluid-insulator transition and graphene}
\label{sec:sit}

Here we review recent new results \cite{m2cft,nernst} 
for electrical and thermal transport in the quantum critical region of a generic CFT in 2+1 dimensions.
Thus we can envisage applications to the superfluid-insulator transition, and have presented scenarios under which
such a framework can be used to interpret measurements of the Nernst effect in the cuprates \cite{nernst,natphys}. 
We have also described a separate set of applications to graphene \cite{graph1,graph2,graph3}: 
while graphene is strictly not a CFT3, the Dirac spectrum of electrons
leads to many similar results, especially in the inelastic collision-dominated regime associated with the quantum critical region.
These results on graphene are reviewed in a separate paper \cite{graphrev}, where explicit microscopic computations are also discussed.

The results presented below were obtained in two separate computations, associated with the methods described in Sections~\ref{sec:qct}
and~\ref{sec:ads}:
\begin{itemize}
\item 
We solved the classical Einstein-Maxwell equations for linearized fluctuations about the metric of a dyonic black hole
in a space which becomes AdS as $u \rightarrow \infty$. The results were used to obtain correlators of a CFT3 using Eq.~(\ref{dual}).
We have no detailed knowledge of the strongly-coupled quantum gravity theory which is dual to the CFT3 describing
the superfluid-insulator transition in condensed matter systems. Nevertheless, it is a reasonable postulate that its low
energy effective field theory essentially captured by the Einstein-Maxwell theory. Armed with this postulate, we can then obtain a powerful set
of results for CFT3s. There is some loss of quantitative accuracy compared to the solvable supersymmetric theory noted above: 
we no longer have the precise quantitative values of $L/\ell_P$ and a few other couplings in the effective Einstein-Maxwell theory,
and so cannot compute the values of universal numerical constants like $\Phi_{\sigma,\eta,s}$ of the CFT3. 
\item
With the picture of relaxation to local equilibrium at frequencies $\hbar \omega \ll k_B T$ developed in Ref.~\cite{damless},
we postulate that the equations of relativistic magnetohydrodynamics should describe the low frequency transport.
The basic principles involved in such a hydrodynamic computation go back to the nineteenth century: conservation
of energy, momentum, and charge, and the constraint of the positivity of entropy production. Nevertheless,
the required results were not obtained until our recent work \cite{nernst}: the general case of a CFT3 in the presence of
a chemical potential, magnetic field, and small density of impurities is very intricate, and the guidance
provided by the dual gravity formulation was very helpful to us. In this approach, we do not have quantitative
knowledge of a few transport co-efficients, and this is complementary to our ignorance of the effective
couplings in the Einstein-Maxwell theory noted above.
\end{itemize}
In the end, we obtained complete agreement between these two independent computations. This agreement demonstrates that
the assumption of a low energy Einstein-Maxwell effective field theory for a strongly coupled theory of quantum gravity
is equivalent to the assumption of hydrodynamic transport for $\hbar \omega \ll k_B T$ in a strongly coupled CFT3.
See also Ref.~\cite{shiraz}, and references therein.

Finally, we turn to our explicit results for quantum critical transport with $\hbar \omega \ll k_B T$. 

First, consider adding a chemical potential, $\mu$,  to the CFT3. This will induce a non-zero
number density of carriers $\rho$. The value of $\rho$ is defined so that the total charge density associated with
$\rho$ is $e^{\ast} \rho$. Note that we can choose $e^{\ast}$ at our convenience, and there is no implicit assumption
that the elementary excitations of the CFT3 carry charge $e^{\ast}$. Then the electrical conductivity at a frequency $\omega$ is 
\begin{equation}
\sigma (\omega) = \frac{e^{\ast 2}}{h} \Phi_\sigma + \frac{e^{\ast 2} \rho^2 v^2}{(\varepsilon + P)} \frac{1}{(- i \omega + 1/\tau_{\rm imp})}
\label{sw}
\end{equation}
In this section, we are again using the symbol $v$ to denote the characteristic velocity of the CFT3 because we will need $c$
for the physical velocity of light below. Here $\varepsilon$ is the energy density and $P$ is the pressure of the CFT3.
We have assumed a small density of impurities which lead to a momentum relaxation time $\tau_{\rm imp}$ \cite{nernst,hh}.
In general, $\Phi_\sigma$, $\rho$, $\varepsilon$, $P$, and $1/\tau_{\rm imp}$ will be functions of $\mu/k_B T$
which cannot be computed by hydrodynamic considerations alone. However, apart from $\Phi_\sigma$, these quantities
are usually amenable to direct perturbative computations in the CFT3, or by quantum Monte Carlo studies.
The physical interpretation of Eq.~(\ref{sw}) should be evident: adding a charge density $\rho$ leads to an 
additional Drude-like contribution to the conductivity. This extra current cannot be relaxed by collisions between the 
unequal density of particle and hole excitations, and so requires an impurity relaxation mechanism to yield a finite
conductivity in the d.c. limit.

Now consider thermal transport in a CFT3 with a non-zero $\mu$. The d.c. thermal conductivity, $\kappa$, is given by
\begin{equation}
\kappa = \Phi_\sigma \left( \frac{k_B^2 T}{h} \right) \left( \frac{\varepsilon + P}{k_B T \rho}
\right)^2 , \label{kapparho}
\end{equation}
in the absence of impurity scattering, $1/\tau_{\rm imp} \rightarrow 0$. This is a Wiedemann-Franz-like relation, connecting the thermal conductivity to the electrical conductivity
in the $\mu=0$ CFT. Note that $\kappa$ diverges as $\rho \rightarrow 0$, 
and so the thermal conductivity of the $\mu=0$ CFT is infinite.

Next, turn on a small magnetic field $B$; we assume that $B$ is small enough
that the spacing between the Landau levels is not as large as $k_B T$. The case
of large Landau level spacing is also experimentally important, but cannot be addressed
by the present analysis. Initially, consider the case $\mu=0$. In this case, the result
Eq.~(\ref{kapparho}) for the thermal conductivity is replaced by
\begin{equation}
\kappa = \frac{1}{\Phi_\sigma} 
\left( \frac{k_B^2 T}{h} \right) \left( \frac{\varepsilon + P}{k_B T B/(hc/e^\ast)}
\right)^2 \label{kappaB}
\end{equation}
also in the absence of impurity scattering, $1/\tau_{\rm imp} \rightarrow 0$.
This result relates $\kappa$ to the electrical {\em resistance} at criticality,
and so can be viewed as Wiedemann-Franz-like relation for the vortices.
A similar $1/B^2$ dependence of $\kappa$ appeared in the Boltzmann equation
analysis of Ref.~\cite{bgs}, but our more general analysis applies in a wider and distinct regime
\cite{graph3}, and relates the
co-efficient to other observables.

We have obtained a full set of results for the frequency-dependent thermo-electric
response functions at non-zero $B$ and $\mu$. The results are lengthy and we refer
the reader to Ref.~\cite{nernst} for explicit expressions. Here we only note that the characteristic feature  \cite{nernst,sean2} of these results is a new {\em hydrodynamic cyclotron resonance}.
The usual cyclotron resonance occurs at the classical cyclotron frequency, which is independent of the particle density and temperature; 
further, in a Galilean-invariant system this resonance
is not broadened by electron-electron interactions alone, and requires impurities
for non-zero damping. The situtation for our hydrodynamic resonance is very different.
It occurs in a collision-dominated regime, and its frequency depends on 
the density and temperature: the explicit expression for the resonance frequency is
\begin{equation}
\omega_c = \frac{e^\ast B \rho v^2}{c (\varepsilon + P)}.
\end{equation}
Further, the cyclotron resonance involves particle and hole excitations moving
in opposite directions, and collisions between them can damp the resonance
even in the absence of impurities. Our expression for this intrinsic damping frequency is \cite{nernst,sean2}
\begin{equation}
\gamma = \frac{e^{\ast 2}}{h} \Phi_\sigma \frac{ B^2 v^2}{c^2 (\varepsilon + P)},
\end{equation}
relating it to the quantum-critical conductivity as a measure of collisions between 
counter-propagating particles and holes.
We refer the reader to a separate discussion \cite{graph1} of the experimental conditions under
which this hydrodynamic cyclotron resonance may be observed.

\section{Recent developments}
\label{sec:recent}

In this concluding section we mention a number of new research directions that have
emerged since our work in Refs.~\cite{m2cft,nernst}. These works are generally
aimed in two directions: to move out of the quantum critical regime of Fig.~\ref{qcfig}
into the low-temperature regimes, and to address a wider class of quantum critical
points of interest in condensed matter physics.

We see from Fig.~\ref{qcfig} that one possible fate of the quantum critical system at low $T$
is to undergo a phase transition to a superfluid state. Gubser and collaborators \cite{gubser1,gubser1a,gubser2,gubser3},
Hartnoll, Herzog, and Horowitz \cite{hhh1,hhh2,sean3,hhh3} and others \cite{o1,o2,o3,o4,o5} have argued
that the AdS4 dual of 
such a transition would involve condensation of a charged scalar field in the background an electrically
charged black hole. Their solutions of such models have yielded fascinating results for the conductivity
of the superfluid state, with intriguing connections to, and differences from, conventional weak coupling
results in the condensed matter literature.

Lee \cite{sslee} has recently studied the fermionic counterpart, by examining a non-zero density
of charged fermions in the background of an electrically charged black hole. The Green's functions
of the boundary theory appear to display an unusual non-Fermi liquid singularity over a ``Fermi ball''.

Extensions which break parity and time-reversal, of relevance to the quantum Hall effect, 
have also been studied \cite{qhe1,qhe2}.

A separate set of developments involve looking at quantum critical theories whose symmetry group
is not the relativistic conformal group considered in the present paper. The most prominent
example of such a quantum critical point is the three-dimensional non-relativistic Fermi
gas near a Feshbach resonance. The phase diagram of this model is efficiently described
by a scale-invariant RG fixed point at zero density \cite{nikolic}; this fixed point 
was shown \cite{son1} to have a non-relativistic conformal symmetry (the Schr\"odinger group).
A number of investigators \cite{son2,uf1,uf2,uf3,uf3a,uf4,uf5,uf5a,uf5b,uf5c,uf5d,uf5e,uf6} have realized the Schr\"odinger group
as the group of isometries of a gravity theory, and obtained interesting information on the low
temperature properties of the boundary critical theory dual to this gravity theory.

Finally, we mention studies \cite{kachru,uf6,pai,maeda} of gravity duals of quantum critical theories with a general
value of the dynamic critical exponent, $z$.

\section*{Acknowledgements}

We thank Sean Hartnoll for a number of valuable clarifications, and for helpful
remarks on the manuscript.
This research was supported by the NSF under grant DMR-0757145, by the FQXi foundation, and
by the Swiss National 
Fund for ScientiÞc Research under grant PA002-113151.


\section*{References}
\begin{thebibliography}{99}

\bibitem{ssbook} S.~Sachdev, {\em Quantum Phase Transitions}, Cambridge University
Press, Cambridge (1999).

\bibitem{giam} T.~Giamarchi, {\em Quantum Physics in One Dimension}, Oxford University
Press, Oxford (2004).
 
\bibitem{damless} K.~Damle and S.~Sachdev, Phys.\ Rev.\ B {\bf 56}, 8714 (1997), arXiv:cond-mat/9705206.

\bibitem{m2cft} C.~P.~Herzog, P.~K.~Kovtun. S.~Sachdev,
and D.~T.~Son, Phys. Rev. D {\bf 75}, 085020 (2007), arXiv:hep-th/0701036.

\bibitem{sondhi1} A.~G.~Green and S.~L.~Sondhi,
Phys.\ Rev.\ Lett.\ {\bf 95}, 267001 (2005), arXiv:cond-mat/0501758.

\bibitem{bgs}  M.~J.~Bhaseen, A.~G.~Green, and S.~L.~Sondhi,
Phys. Rev. Lett. {\bf 98}, 166801 (2007), arXiv:cond-mat/0610687.

\bibitem{MAGOO}
  O.~Aharony, S.~Gubser, J.~Maldacena, H.~Ooguri and Y.~Oz,
  Phys.\ Rept.\ {\bf 323}, 183 (2000), arXiv:hep-th/9905111.
  
\bibitem{nernst}  S.~A.~Hartnoll, P.~K.~Kovtun, M.~M\"uller,
and S.~Sachdev, Phys. Rev. B {\bf 76}, 144502 (2007), arXiv:0706.3215.

\bibitem{fwgf} M.~P.~A.~Fisher, P.~B.~Weichman, G.~Grinstein, and D.~S.~Fisher,
Phys.\ Rev.\ B {\bf 40}, 546 (1989).

\bibitem{hod} S. Hod, Phys. Rev. D {\bf 75} 064013 (2007), arXiv:gr-qc/0611004; 
Class. Quantum Grav. {\bf 24}, 4235 (2007), arXiv:0705.2306.

\bibitem{pesci} A.~Pesci, arXiv:0807.0300.

\bibitem{pkk1}  P.~K.~Kovtun and A.~Ritz, Phys. Rev. D {\bf 78}, 066009 (2008), arXiv:0806.0110.

\bibitem{kss} P.~K.~Kovtun, D.~T.~Son, and A.~Starinets, Phys. Rev. Lett.  {\bf 94}, 11601 (2005),
arXiv:hep-th/0405231.

\bibitem{pkk2} P.~K.~Kovtun and A.~Ritz, Phys. Rev. Lett. {\bf 100}, 171606 (2008), arXiv:0801.2785.

\bibitem{sspoly} S.~Sachdev, Physics Letters B {\bf 309}, 285 (1993), arXiv:hep-th/9305131.

\bibitem{beck}  J.~D.~Bekenstein,
Phys. Rev. D {\bf 7}, 2333 (1973).

\bibitem{hawking} S.~W.~Hawking, Comm. Math. Phys. {\bf 43}, 199 (1975).

\bibitem{hhh1} S.~A.~Hartnoll, C.~P.~Herzog, and G.~T.~Horowitz, arXiv:0803.3295.

\bibitem{sean1} S.~A.~Hartnoll and P.~K.~Kovtun, Phys. Rev. D {\bf 76}, 066001 (2007), arXiv:0704.1160.

\bibitem{gubser0} S.~S.~Gubser, I.~R.~Klebanov, and A.~M.~Polyakov, Phys. Lett. B {\bf 428}, 105 (1998),
arXiv:hep-th/9802109.

\bibitem{witten} E.~Witten, Adv. Theor. Math. Phys. {\bf 2}, 253 (1998), arXiv:hep-th/9802150.

\bibitem{malda} J.~M.~Maldacena, JHEP {\bf 0304}, 021 (2003), arXiv:hep-th/0106112.

\bibitem{sonherzog} D.~T.~Son and A.~O.~Starinets, JHEP {\bf 0209}, 042 (2002), arXiv:hep-th/0205051;
C.~P.~Herzog and D.~T.~Son, JHEP {\bf 0303}, 046 (2003), arXiv:hep-th/0212072.

\bibitem{kleb} I.~R.~Klebanov and A.~A.~Tseytlin, Nucl. 
Phys. B {\bf 475} 164 (1996), arXiv:hep-th/9604089.

\bibitem{chris} C.~P.~Herzog,  JHEP {\bf 0212}, 026 (2002), arXiv:hep-th/0210126.

\bibitem{natphys} S.~Sachdev, Nature Physics {\bf 4}, 173 (2008), arXiv:0711.3015.

\bibitem{graph1} M. M{\"u}ller and S. Sachdev,
{ Phys. Rev. B} {\bf 78}, 115419 (2008), arXiv:0801.2970.

\bibitem{graph2}
L. Fritz, J. Schmalian, M. M{\"u}ller, and S. Sachdev,
{ Phys. Rev. B} {\bf 78}, 085416 (2008), arXiv:0802.4289.

\bibitem{graph3}
M. M\"uller, L. Fritz, and S. Sachdev,
{ Phys. Rev. B} {\bf 78}, 115406 (2008), arXiv:0805.1413.

\bibitem{graphrev} M. M{\"u}ller, L. Fritz, S. Sachdev, and J. Schmalian, to appear.

\bibitem{shiraz} S.~Bhattacharyya, S.~Minwalla, and S.~R.~Wadia, arXiv:0810.1545.

\bibitem{hh} S.~A.~Hartnoll and C.~P.~Herzog, Phys. Rev. D {\bf 77}, 106009 (2008), arXiv:0801.1693.

\bibitem{sean2} 
  S.~A.~Hartnoll and C.~P.~Herzog,
  Phys.\ Rev.\  D {\bf 76}, 106012 (2007), arXiv:0706.3228.


\bibitem{gubser1} S.~S.~Gubser, arXiv:0801.2977.

\bibitem{gubser1a} S.~S.~Gubser, arXiv:0803.3483.

\bibitem{gubser2} S.~S.~Gubser and S.~S.~Pufu, arXiv:0805.2960.

\bibitem{gubser3} S.~S.~Gubser and F.~D.~Rocha, arXiv:0807.1737.
 
\bibitem{hhh2} G.~T.~Horowitz and M.~M.~Roberts, arXiv:0803.3295.

\bibitem{sean3} M.~M.~Roberts and S.~A.~Hartnoll, arXiv:0805.3898.
 
\bibitem{hhh3} S.~A.~Hartnoll, C.~P.~Herzog, and G.~T.~Horowitz, arXiv:0810.1563.

\bibitem{o1} E.~Nakano and W.-Y.~Wen, Phys. Rev. D {\bf 78}, 046004 (2008), arXiv:0804.3180.

\bibitem{o2} T.~Albash and C.~V.~Johnson, JHEP {\bf 09}, 121 (2008), arXiv:0804.3466.

\bibitem{o3} W.-Y.~Wen, arXiv:0805.1550.

\bibitem{o4} K.~Maeda and T.~Okamura, arXiv:0809.3079.

\bibitem{o5} C.~P.~Herzog, P.~K.~Kovtun, and D.~T.~Son, arXiv:0809.4870.

\bibitem{sslee} S.-S. Lee, arXiv:0809.3402.

\bibitem{qhe1} E.~Keski-Vakkuri and P.~Kraus, arXiv:0805.4643.

\bibitem{qhe2} J.~L.~Davis, P.~Kraus, and A.~Shah, arXiv:0809.1876.

\bibitem{nikolic} P. Nikoli\'c and S. Sachdev,  Phys. Rev. A 
{\bf 75}, 033608 (2007), arXiv:cond-mat/0609106.

\bibitem{son1} Y.~Nishida and D.~T.~Son, Phys. Rev. D {\bf 76}, 086004 (2007), arXiv:0706.3746.

\bibitem{son2} D.~T.~Son, arXiv:0804.3972.

\bibitem{uf1} K.~Balasubramaniam and J.~McGreevy, arXiv:0804.4053.

\bibitem{uf2} W.~D.~Goldberger, arXiv:0806.2867. 

\bibitem{uf3} J.~L.~F.~Barb\'on and C.~A.~Fuertes, arXiv:0806.3244.

\bibitem{uf3a} C.~P.~Herzog, M.~Rangamani, and S.~F.~Ross, arXiv:0807.1099.

\bibitem{uf4} J.~Maldacena, D.~Martelli, and Y.~Tachikawa, arXiv:0807.1100.

\bibitem{uf5} A.~Adams, K.~Balasubramanian, and J.~McGreevy, arXiv:0807.1111. 

\bibitem{uf5a} D.~Minic and M.~Pleimling, arXiv:0807.3665.

\bibitem{uf5b} J.-W.~Chen and W.-Y.~Wen, arXiv:0808.0399.

\bibitem{uf5c} P.~K.~Kovtun and D.~Nickel, arXiv:0809.2020.

\bibitem{uf5d} C.~Duval, M.~Hassaine, and P.~A.~Horvathy, arXiv:0809.3128.

\bibitem{uf5e} D.~Yamada, arXiv:0809.4928.

\bibitem{uf6} S.~A.~Hartnoll and K.~Yoshida, arXiv:0810.0298.

\bibitem{kachru} S.~Kachru, X.~Liu, and M.~Mulligan, arXiv:0808.1725.

\bibitem{pai} S.~Pai, arXiv:0809.1756.

\bibitem{maeda} K.~Maeda, M.~Natsuume, and T.~Okamura, arXiv:0809.4074.


\end{thebibliography}
\end{document}